\documentclass[conference, 9pt]{IEEEtran}
\usepackage{spconfa4,graphicx}
\usepackage{graphicx}
\usepackage{multirow}
\usepackage{amsmath,amssymb,amsfonts}
\usepackage{amsthm}
\usepackage{mathrsfs}
\usepackage[figuresright]{rotating}
\usepackage{xcolor}
\usepackage{textcomp}
\usepackage{manyfoot}
\usepackage{booktabs}
\usepackage{algorithm}
\usepackage{algorithmicx}
\usepackage{algpseudocode}
\usepackage{program}
\usepackage{listings}
\usepackage{cite}

\begin{document}

\title{Adaptive Diagonal Loading using Krylov Subspaces for Robust Beamforming}

\name{Manan Mittal $^1$, Ryan M. Corey $^{2}$, John R. Buck $^3$, Andrew C. Singer $^1$ \thanks{Funded in part by US Navy, Office of Naval Research under award N00014-23-1-2133}}
\address{Stony Brook University$^1$, University of Illinois Chicago$^2$, University of Massachusetts Dartmouth$^3$}

\maketitle

\begin{abstract}
Reliable adaptive beamforming is critical for large microphone arrays operating in highly dynamic acoustic environments. In scenarios characterized by fast-moving talkers and interferers, the available sample support for estimating the spatial correlation matrix is often snapshot-deficient. This deficiency degrades the White Noise Gain (WNG), leading to severe target signal cancellation. To ensure stable and robust beamforming, we previously proposed an adaptive diagonal loading method that leverages the Kantorovich inequality to guarantee the WNG remains strictly within specified bounds. However, accurately determining the smallest necessary loading level requires calculating the extreme eigenvalues of the spatial correlation matrix, a computationally expensive $\mathcal{O}(M^3)$ operation for large arrays. In this paper, we introduce a highly efficient $\mathcal{O}(kM^2)$ estimation technique using Lanczos iterations to build a small Krylov subspace. By projecting the correlation matrix onto a tridiagonal matrix of dimension $k \ll M$, we extract Ritz values that rapidly converge to the exact extreme eigenvalues. Our evaluations demonstrate that this Lanczos-accelerated approach achieves performance identical to exact Eigenvalue Decomposition (EVD), ensuring optimal interference suppression and strict WNG adherence at a fraction of the computational cost.
\end{abstract}

\section{Introduction}
Adaptive beamforming techniques, such as the Minimum Power Distortionless Response (MPDR) and Minimum Variance Distortionless Response (MVDR) \cite{capon}, achieve high spatial resolution by adapting their spatial filter weights to the second-order statistics of the received acoustic data. However, ensuring the robustness of these adaptive beamformers remains a significant challenge when deploying large microphone arrays in dynamic environments characterized by fast-moving sources.

The fundamental vulnerability lies in the reliance on the sample Spatial Correlation Matrix (SCM) \cite{gannot2017consolidated}. To accurately track a fast-moving scene, the observation window must be kept exceedingly short. When the number of available snapshots is less than the number of microphone elements, the SCM becomes poorly conditioned or mathematically rank-deficient. Sample matrix inversion under this snapshot deficiency causes the spatial weights to become highly erratic, resulting in extreme sensitivity to spatially uncorrelated noise, a dramatic collapse in the White Noise Gain (WNG), and severe cancellation of the target signal \cite{cox1987robust}.

Diagonal Loading (DL) is the classical remedy to mitigate SCM ill-conditioning by artificially inflating the spatial noise floor \cite{van2002optimum}. While standard DL is ubiquitous, selecting the optimal loading parameter $\mu$ is historically an ad-hoc process. We have previously demonstrated a dynamic, closed-form adaptive diagonal loading method that deterministically guarantees the WNG stays within specified bounds by exploiting the strict mathematical relationship between the array's WNG and the condition number of the SCM via the Kantorovich inequality \cite{mittal2026adaptive}.

Because computing the exact eigenvalues of the SCM to determine the necessary DL at every time step requires an exact Eigenvalue Decomposition (EVD) scaling at $\mathcal{O}(M^3)$, it is computationally prohibitive for massive arrays operating at high sample rates. Relaxed bounding methods, such as the Gershgorin Circle Theorem or Trace-based bounds, offer lower complexity but overestimate the required loading, thereby unnecessarily penalizing the beamformer's degrees of freedom. 

To bridge this gap, we extend our adaptive loading framework by introducing a Krylov subspace method via the Lanczos algorithm. A Krylov subspace is iteratively constructed by repeatedly multiplying a matrix by a starting vector. Because repeated multiplication naturally amplifies the influence of the largest and smallest eigenvalues, the resulting subspace is highly biased toward the matrix's extreme eigenspaces. The Lanczos algorithm leverages this property to mathematically project the massive $M \times M$ spatial correlation matrix onto a much smaller, $k \times k$ tridiagonal matrix, where $k \ll M$. The extreme eigenvalues of this tiny projected matrix—known as Ritz values—rapidly converge to the true extreme eigenvalues of the original correlation matrix. By relying on just $k$ matrix-vector multiplications rather than a full matrix decomposition, this approach yields an highly efficient $\mathcal{O}(kM^2)$ algorithm. It perfectly matches the precision of an exact EVD solver, extracting the exact bounding parameters needed for optimal diagonal loading without the prohibitive computational overhead.

\section{Signal Model}
We consider a room acoustic environment capturing a group of $J$ active sound sources using an array of $M$ microphones. Following the narrowband multiplicative assumption in the Short-Time Fourier Transform (STFT) domain, the vectorized array signal model at frame $i$ is written as:
\begin{equation}
    \mathbf{y}[i] = \mathbf{H}\mathbf{s}[i] + \mathbf{v}[i]
\end{equation}
where $\mathbf{y}[i] \in \mathbb{C}^{M \times 1}$, $\mathbf{s}[i] \in \mathbb{C}^{J \times 1}$, and $\mathbf{H} \in \mathbb{C}^{M \times J}$ is the matrix of acoustic transfer functions. For a target source of interest, we define the relative steering vector $\mathbf{d} \in \mathbb{C}^{M \times 1}$, normalized such that $\mathbf{d}^H \mathbf{d} = M$.

The MPDR beamformer seeks a weight vector $\mathbf{w}[i] \in \mathbb{C}^{M \times 1}$ that minimizes output power while maintaining a distortionless target response:
\begin{equation}
    \min_{\mathbf{w}} \mathbf{w}^H \mathbf{R}_y \mathbf{w} \quad \text{s.t.} \quad \mathbf{w}^H \mathbf{d} = 1
\end{equation}
where $\mathbf{R}_y = \mathbb{E}[\mathbf{y}\mathbf{y}^H]$ is the theoretical SCM. In practice, the true SCM is unknown and must be approximated via a short sliding window of length $L$:
\begin{equation}
    \hat{\mathbf{R}}_y[i] = \frac{1}{L} \sum_{l=0}^{L-1} \mathbf{y}[i-l]\mathbf{y}^H[i-l]
\end{equation}
When $L < M$, $\hat{\mathbf{R}}_y[i]$ is rank-deficient, heavily amplifying uncorrelated noise during matrix inversion.

\section{Proposed Method}

\subsection{WNG Bounds via Kantorovich Limits}
The robustness of an adaptive beamformer to uncorrelated noise is quantified by its White Noise Gain (WNG), defined as:
\begin{equation}
    W = \frac{1}{\mathbf{w}^H \mathbf{w}}
\end{equation}
where the equality holds due to the distortionless constraint $\mathbf{w}^H \mathbf{d} = 1$. To strictly bound the WNG such that $W \geq W_{\min}$, we leverage the Kantorovich inequality \cite{kantorovich1948functional}. For any Hermitian positive-definite matrix $\mathbf{R}_y$ with condition number $\kappa = \lambda_{\max} / \lambda_{\min}$, the inequality yields the strict relationship:
\begin{equation}
    \frac{W}{M} \geq \frac{4\kappa}{(\kappa+1)^2}
\end{equation}
Let $A_G = M / W_{\min}$ represent the strict array gain limit. Solving this inequality for the maximum allowable condition number $\kappa_{\max}$ yields:
\begin{equation}
    \kappa_{\max} = (2A_G - 1) + 2\sqrt{A_G(A_G - 1)}
\end{equation}
By actively limiting the condition number of the estimated SCM to $\kappa_{\max}$, we strictly control and guarantee the minimum WNG.

\subsection{Adaptive Diagonal Loading Estimation}
To restrict the SCM's condition number, we apply a dynamic diagonal loading factor $\mu[i]$ at every frame:
\begin{equation}
    \mathbf{Q}[i] = \hat{\mathbf{R}}_y[i] + \mu[i] \mathbf{I}
\end{equation}
The eigenvalues of the loaded matrix $\mathbf{Q}[i]$ shift by $\mu[i]$. To satisfy the condition limit $\kappa_{loaded} = (\lambda_{\max} + \mu[i]) / (\lambda_{\min} + \mu[i]) \leq \kappa_{\max}$, we solve for the exact required multiplier:
\begin{equation}
    \mu[i] = \max \left(0, \frac{\lambda_{\max} - \kappa_{\max}\lambda_{\min}}{\kappa_{\max} - 1} \right)
\end{equation}
This ensures we apply the absolute minimum loading necessary, preserving the beamformer's adaptive degrees of freedom for interference cancellation.

\subsection{Krylov Subspace Estimation via Lanczos Iteration}
Calculating $\mu[i]$ requires knowledge of the extreme eigenvalues ($\lambda_{\max}$ and $\lambda_{\min}$) of $\mathbf{Q}[i]$. Because exact EVD is computationally prohibitive ($\mathcal{O}(M^3)$), we propose utilizing the Lanczos algorithm to iteratively build a small orthogonal basis for the Krylov subspace:
\begin{equation}
    \mathcal{K}_k(\mathbf{Q}, \mathbf{v}_1) = \text{span}\{\mathbf{v}_1, \mathbf{Q}\mathbf{v}_1, \mathbf{Q}^2\mathbf{v}_1, \dots, \mathbf{Q}^{k-1}\mathbf{v}_1\}
\end{equation}
where $k \ll M$ is the number of iterations. We initialize the algorithm with a normalized uniform vector $\mathbf{v}_1 = \mathbf{1}/\sqrt{M}$. Note, this a design choice and doesn't restrict the method from being used for arbitrary array geometries. For $j = 1, \dots, k$, we perform the following recursion:
\begin{align}
    \mathbf{w}_j &= \mathbf{Q}\mathbf{v}_j \\
    \mathbf{w}_j &= \mathbf{w}_j - \beta_{j-1}\mathbf{v}_{j-1} \quad (\text{if } j > 1) \\
    \alpha_j &= \text{Re}(\mathbf{v}_j^H \mathbf{w}_j) \\
    \mathbf{w}_j &= \mathbf{w}_j - \alpha_j \mathbf{v}_j \\
    \beta_j &= \|\mathbf{w}_j\|_2 \\
    \mathbf{v}_{j+1} &= \mathbf{w}_j / \beta_j
\end{align}
This iterative process projects the massive $M \times M$ matrix $\mathbf{Q}$ onto a highly compact $k \times k$ symmetric tridiagonal matrix $\mathbf{T}_k$:
\begin{equation}
    \mathbf{T}_k = \begin{bmatrix}
    \alpha_1 & \beta_1 & 0 & \dots & 0 \\
    \beta_1 & \alpha_2 & \beta_2 & \dots & 0 \\
    0 & \beta_2 & \alpha_3 & \dots & 0 \\
    \vdots & \vdots & \vdots & \ddots & \beta_{k-1} \\
    0 & 0 & 0 & \beta_{k-1} & \alpha_k
    \end{bmatrix}
\end{equation}

By the properties of Krylov subspace methods, the extreme eigenvalues of $\mathbf{T}_k$ (known as Ritz values) rapidly converge to the true extreme eigenvalues of the full matrix $\mathbf{Q}$. Thus, we make the highly accurate approximations:
\begin{equation}
    \lambda_{\max}(\mathbf{Q}) \approx \lambda_{\max}(\mathbf{T}_k), \quad \lambda_{\min}(\mathbf{Q}) \approx \lambda_{\min}(\mathbf{T}_k)
\end{equation}

Because the dimension $k$ is extremely small (e.g., $k=4$), calculating the exact EVD of $\mathbf{T}_k$ is computationally trivial, requiring roughly $\mathcal{O}(k^3)$ operations. The dominant computational cost of this method lies entirely in the $k$ matrix-vector multiplications ($\mathbf{Q}\mathbf{v}_j$) required to build the subspace, yielding an overall computational complexity of $\mathcal{O}(kM^2)$. 

\section{Simulations \& Discussion}
We evaluate the proposed adaptive diagonal loading strategy using a simulated uniform linear array (ULA) consisting of $M=15$ microphones with half-wavelength spacing at a center frequency of $f_0 = 1000$ Hz. To rigorously test the tracking capabilities and robustness of the algorithms, we simulate a highly dynamic ``birth-death'' spatial interference scenario over $T=20000$ snapshots for $200$ Monte Carlo trials. In this scenario, up to two statistically independent interferers randomly appear, remain active for a duration, and disappear. 
\begin{figure}[t]
    \centering
    \includegraphics[width=\columnwidth]{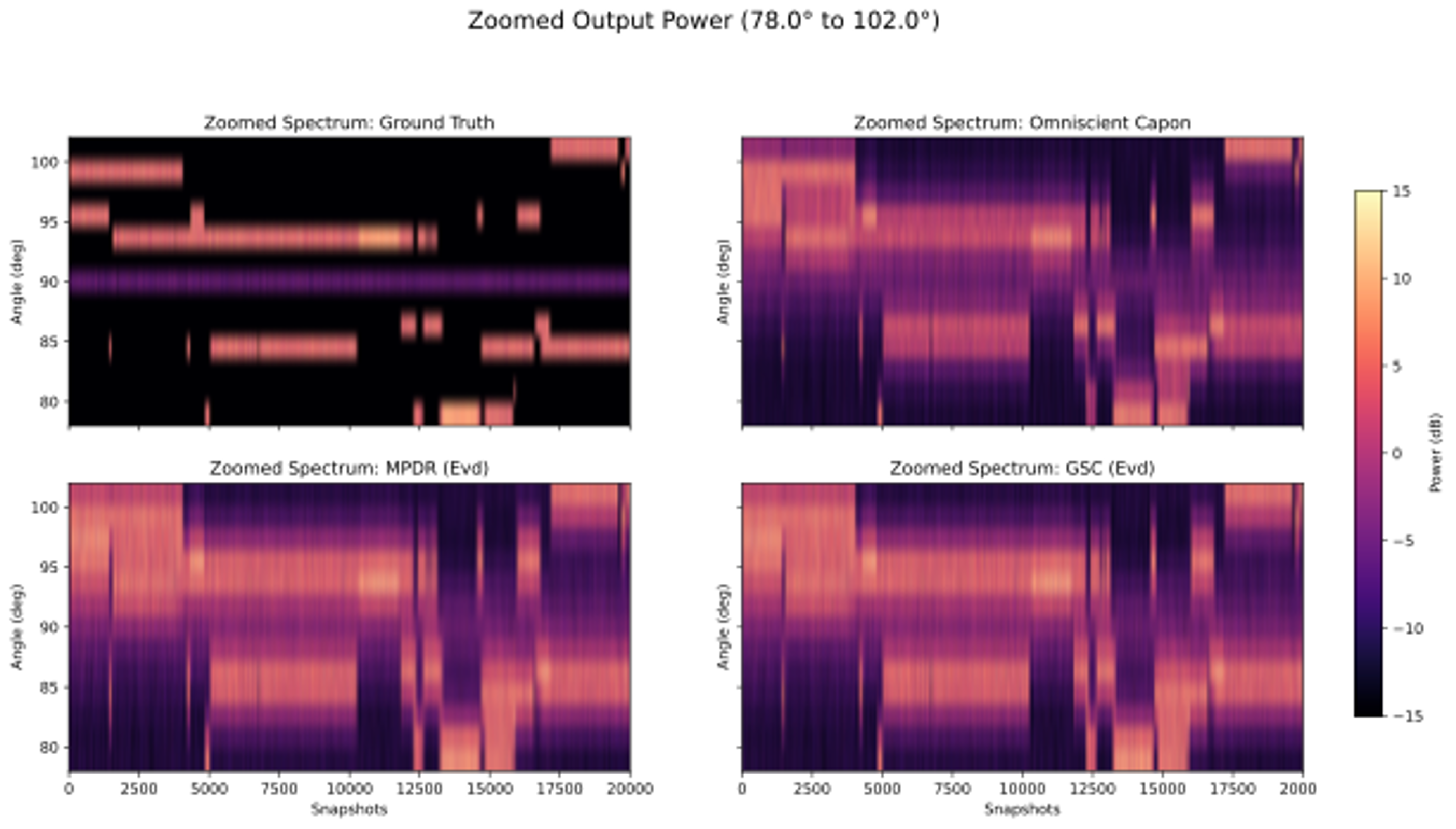} 
    \caption{The scanned response for the ground truth, Omniscient, EVD based diagonal loading, and proposed method. The scanned responses demonstrate that the proposed method sacrifices no performance but is computationally more tractable.}
    \label{fig:scanned_response}
\end{figure}

To prevent trivial interference scenarios or impossible target separation, the interferers are strictly confined to an angular grid where the target's normalized quiescent beampattern response falls between $-13$ dB and $-3$ dB. This may be typical in cocktail party scenario where multiple closely spaced talkers may need to be separated. The dynamic interferers are generated with an Interference-to-Noise Ratio (INR) of $7$ dB. The target signal is fixed at broadside ($90^\circ$) with a Signal-to-Noise Ratio (SNR) of $-5$ dB. To induce snapshot deficiency, the sample Spatial Correlation Matrix (SCM) is tracked using a sliding rectangular window of $L = 37$ snapshots ($L \approx 2.5M$). For an array of $M=15$, the maximum theoretical WNG is $10 \log_{10}(15) \approx 11.76$ dB. To allow for adaptive interference nulling while preventing target cancellation, we define a strict WNG lower bound of $W_{\min} = 10 \log_{10}(M) - 3 \approx 8.76$ dB. The extreme eigenvalues were computed using the Exact EVD ($\mathcal{O}(M^3)$) and the proposed Lanczos method using only $k=4$ iterations. Figure \ref{fig:scanned_response} shows the ground truth spatial response, alongside the omniscient Capon beamformer that has access to the ground truth statistics for one trial. In addition, we include the scanned response of the proposed method and the exact eigenvalue decomposition. 
\begin{figure}[t]
    \centering
    \includegraphics[width=\columnwidth]{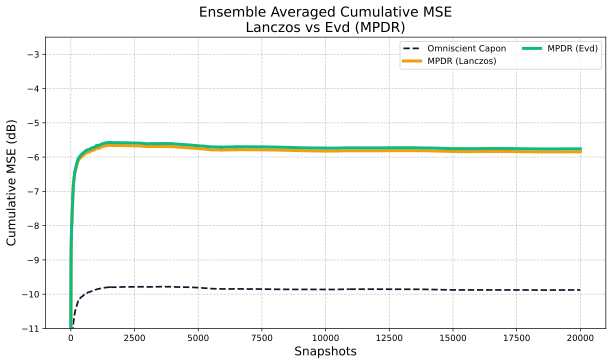} 
    \caption{Ensemble output mean-squared error. The proposed method achieves similar performance with less computation.}
    \label{fig:mse_ensemble}
\end{figure}

\begin{figure}[t]
    \centering
    \includegraphics[width=\columnwidth]{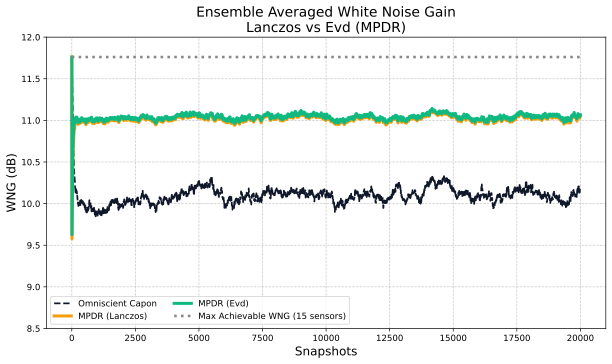} 
    \caption{Ensemble output white noise gain. The proposed method stays within the white noise gain bound of $8.76 db$ and matches the loading level determined by the exact eigenvalue decomposition.}
    \label{fig:wng_ensemble}
\end{figure}

\begin{figure}[t]
    \centering
    \includegraphics[width=\columnwidth]{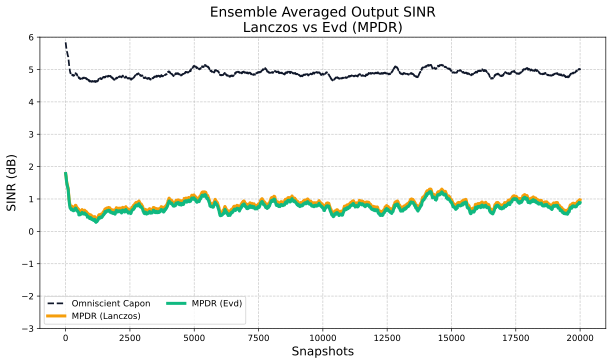} 
    \caption{Ensemble output SINR. The Exact EVD and Lanczos modes compared to the Omniscient baseline. Notably Lanczos performs comparably to the exact EVD.}
    \label{fig:sinr_ensemble}
\end{figure}

\begin{figure}[t]
    \centering
    \includegraphics[width=\columnwidth]{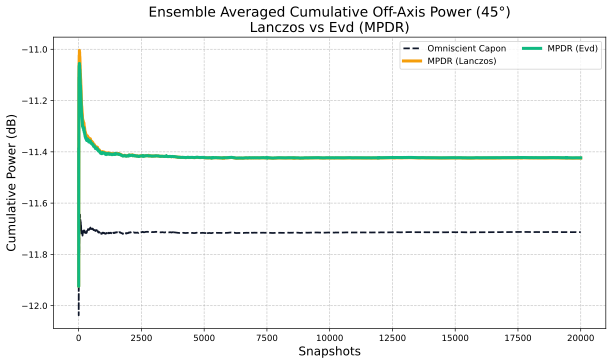} 
    \caption{Output power at 45 degrees. In the absence of a signal this is equivalent to mean-squared error. The propsed method performs comparably to the exact EVD with less compute.}
    \label{fig:off_axis_ensemble}
\end{figure}

Our evaluations demonstrate that the Lanczos approach performs identically to the exact eigenvalue decomposition. Figure \ref{fig:mse_ensemble} compares the mean-squared estimation error of the proposed method and the eigenvalue decomposition. Due to the ability of Krylov subspace methods to efficiently converge to the extreme ends of the matrix spectrum, a mere $k=4$ iterations are sufficient to accurately capture $\lambda_{\max}$ and $\lambda_{\min}$ for the bounds evaluation. Consequently, the computed required loading parameter $\mu_{req}$ under the Lanczos mode is virtually indistinguishable from the optimally derived $\mu$ via exact EVD. The Lanczos-accelerated solver perfectly adheres to the deterministic condition limits, ensuring the array's WNG never drops below the specified $8.76$ dB bound, thereby preventing target cancellation. In Figure \ref{fig:wng_ensemble}, we show that the WNG constraint is met and close to the EVD. The output Signal-to-Interference-plus-Noise Ratio (SINR), Figure \ref{fig:sinr_ensemble}, confirms that the $\mathcal{O}(kM^2)$ Lanczos mode sacrifices zero performance relative to the full $\mathcal{O}(M^3)$ EVD baseline, retaining maximum degrees of freedom for precise and deep interference null steering. Finally, the output power in an off-axis direction, specifically $45 \deg$ where there are no active sources, is shown in Figure \ref{fig:off_axis_ensemble}. In the absence of a source mean-squared error and output power co-incide and the results demonstrate the the proposed method produces equally good results in other directions, despite the iterations being initialized with the broadside steering vector.

\section{Experiments}
\begin{figure}[t]
    \centering
    \includegraphics[width=\columnwidth]{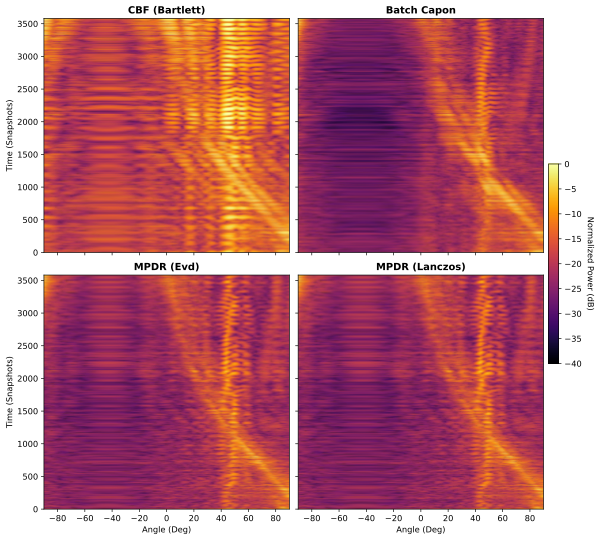} 
    \caption{The bearing time record for the conventional beamformer, batch Capon, EVD based diagonal loading, and proposed method on the SwellEx dataset.}
    \label{fig:swellex_btr}
\end{figure}

\begin{figure}[t]
    \centering
    \includegraphics[width=\columnwidth]{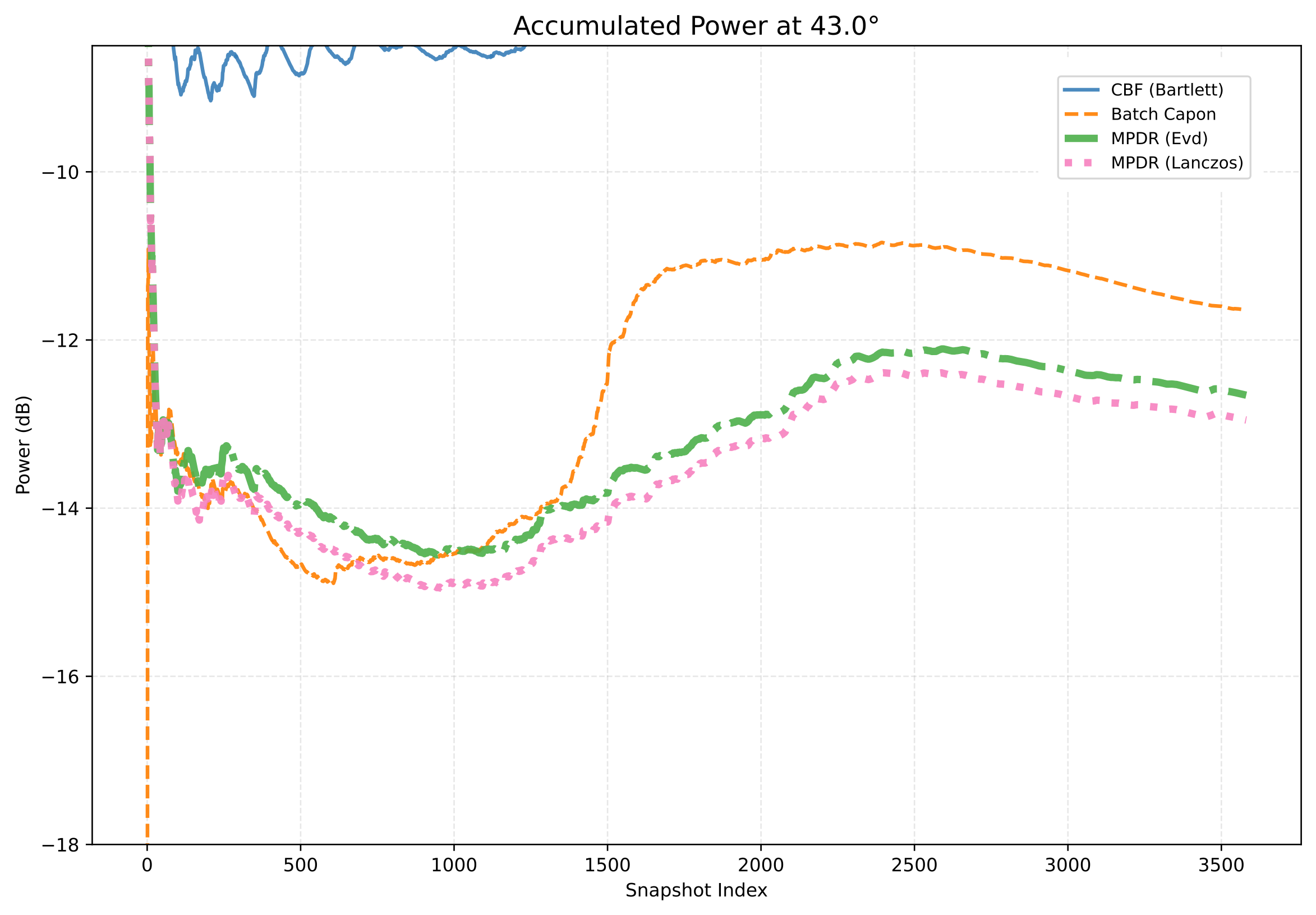} 
    \caption{The output power for the SwellEx dataset at $45 \deg$. This corresponds with the direction of the source with constant bearing relative to the array.}
    \label{fig:swellex_power}
\end{figure}

\begin{figure}[t]
    \centering
    \includegraphics[width=\columnwidth]{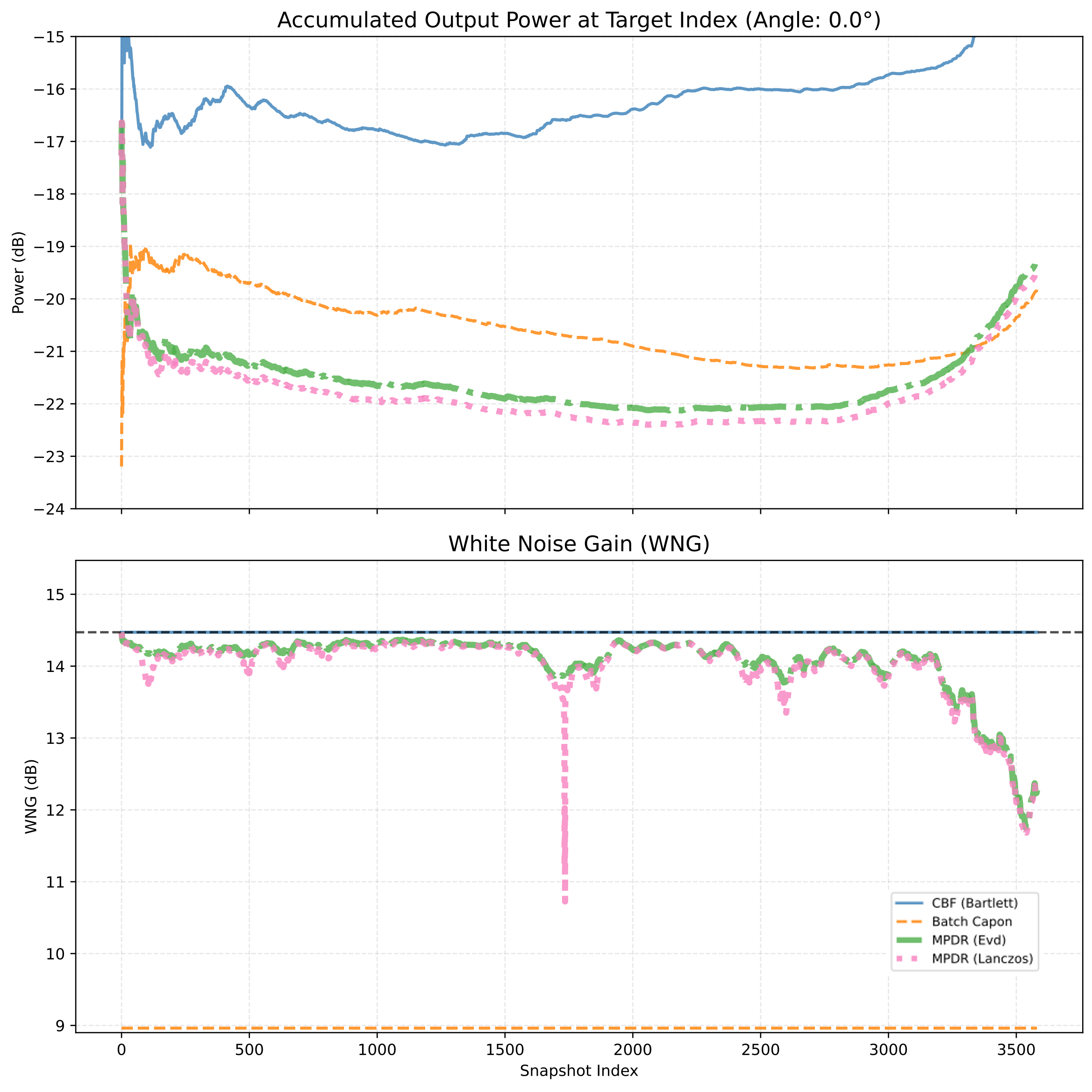} 
    \caption{The off-axis performance of the proposed method. Shown is the accumulated output power at broadside, and the corresponding white noise gain for each of the methods over time.}
    \label{fig:swellex_off_axis}
\end{figure}

The method is validated in a complex, real-world environment using the S59 event from the SwellEx-96 experiment. In this dataset, a source ship tows multiple acoustic projectors through a shallow-water environment ($\approx 200$~m depth) near San Diego, CA. Data from the South Horizontal Line Array are processed Horizontal Line Array (HLA-S), a seafloor array with a $255$~m aperture, using 28 functional hydrophones sampled at $3276.8$~Hz. The scanned response is computed in the horizontal plane ($0^\circ$ elevation) to track the source bearing over time. This scenario rigorously tests adaptability: as the source traverses the multipath environment, the relative geometry and spatial covariance evolve continuously. The proposed lanczos method is compared against the exact eigenvalue decomposition, batch Capon and conventional beamformer. The WNG constraint is set 6dB below the conventional beamformer white noise gain. Figure \ref{fig:swellex_btr} presents the resulting Bearing-Time Records (BTR). In Figure \ref{fig:swellex_power}, we show the accumulated output power at $43 \deg$, corresponding to the source with constant bearing relative to the array, for each of the compared methods. Consistent with previous results, the proposed method performs comparably (marginally better), while requiring lower compute. Figure \ref{fig:swellex_off_axis} demonstrates the off-axis performance of the proposed method and the white noise gain for broadside (directional cosine = $0$). Yet again, the method produces comparable performance to the full EVD while requiring significantly less compute. These results demonstrate that the proposed method is effective in real-world scenarios and can adhere to a strict white noise gain limit as defined by previous methods.

\section{Conclusion}
In this paper, we enhanced our dynamic WNG-constrained beamforming framework by introducing a Krylov subspace eigenvalue estimation method. By employing the Lanczos algorithm, we efficiently project the high-dimensional spatial correlation matrix onto a much smaller tridiagonal subspace to identify the extreme eigenvalues dictating the Kantorovich bounds. This formulation strictly preserves the mathematically guaranteed stability and performance of an exact EVD solver while dropping the algorithmic complexity from $\mathcal{O}(M^3)$ to $\mathcal{O}(kM^2)$. The proposed method is highly scalable and establishes a practical, computationally lightweight pathway for optimally robust beamforming in massive arrays operating at high sample rates.

\bibliographystyle{IEEEbib}
\bibliography{refs}

\end{document}